\documentclass[journal]{rmaa}

\usepackage{paralist}
\usepackage{psfrag,color}
\usepackage[latin1]{inputenc}

\newcommand{\dens}{$n_{\rm e}$}
\newcommand{\temp}{$T_{\rm e}$}
\newcommand*{\TabV}[1]{%
\setlength{\fboxrule}{0pt}\fbox{\hspace*{-\fboxsep}#1\hspace*{-\fboxsep}}}

\title{Physical conditions derived from \ion{O}{2} recombination lines in planetary nebulae  and their 
implications} 
\author{A. Peimbert\altaffilmark{1}, M. Peimbert\altaffilmark{1}, G. Delgado-Inglada\altaffilmark{1},
J. Garc\'ia-Rojas\altaffilmark{2}, and M. Pe\~na\altaffilmark{1}}
\altaffiltext{1}{Instituto de Astronom\'ia, Universidad Nacional Aut\'onoma de M\'exico, Mexico}
\altaffiltext{2}{Instituto de Astrof\'isica de Canarias (IAC), Spain}

\shortauthor{Peimbert et al.}
\shorttitle{Physical conditions derived from \ion{O}{2} RLs in PNe}

\fulladdresses{
\item Antonio Peimbert, Manuel Peimbert, Gloria Delgado-Inglada, and Miriam Pe\~na: 
Instituto de Astronom\'ia, Universidad Nacional Aut\'onoma de M\'exico, Apdo. Postal 70-264. 
M\'exico, 04510 D. F., Mexico (antonio@astro.unam.mx,
 peimbert@astro.unam.mx, gdelgado@astro.unam.mx, miriam@astro.unam.mx).
\item J. Garc\'ia-Rojas: Instituto de Astrof\'isica de Canarias (IAC), 38200, La 
Laguna, Tenerife, Spain; Universidad de La Laguna, Dept. Astrof\'isica, 38206, La Laguna, 
Tenerife, Spain (jogarcia@iac.es).}
  
\listofauthors{A. Peimbert, M. Peimbert, G. Delgado-Inglada, J. Garc\'ia-Rojas, and M. Pe\~na}
\indexauthor{Peimbert, A.}
\indexauthor{Peimbert, M.}
\indexauthor{Delgado-Inglada, G.}
\indexauthor{Garc\'ia-Rojas, J.}
\indexauthor{Pe\~na, M.}

\resumen{A partir de observaciones de alta calidad del multiplete V1 de \ion{O}{2}
estudiamos la densidad y la temperatura de una muestra de nebulosas planetarias.
Encontramos que, en general, las densidades que obtenemos a partir de las l\'ineas 
de \ion{O}{2} {son similares} a las densidades obtenidas a partir de l\'ineas 
prohibidas. Esto implica que no hay evidencia de condensaciones de alta densidad y 
baja temperatura para la mayor\'ia de los objetos de nuestra muestra. Las presiones 
electr\'onicas encontradas en las zonas calientes son semejantes o ligeramente mayores 
que las de las zonas fr\'ias, sugiriendo la presencia de ondas de choque.
Las temperaturas promedio y los valores de $t^2$ obtenidos a partir de l\'ineas de H, 
He y O son similares y consistentes con un medio qu\'imicamente homog\'eneo. Estos 
resultados {sugieren} que las abundancias obtenidas a partir de las l\'ineas de recombinaci\'on 
son las representativas de estos objetos.}

\abstract{Based on high quality observations of multiplet V1 of \ion{O}{2} and
the NLTE atomic computations for \ion{O}{2} we study the density and
temperature of a sample of PNe. We find that, in general, the densities 
derived from recombination lines of \ion{O}{2} {are similar} than the
densities derived from forbidden lines. This implies that the signature for oxygen rich 
clumps of high density and low temperature is absent in most of the objects of our 
sample. Electron pressures derived from the hotter zones are similar or slightly larger 
than those derived from the colder zones, suggesting the presence of shock waves. 
The average temperatures and $t^2$ values derived from
H, He and O lines are similar and consistent with chemical homogeneity.
These results {suggest that the abundances of these objects are the ones derived 
from recombination lines.}}

\addkeyword{planetary nebulae: abundances}
\addkeyword{planetary nebulae: physical conditions}
\addkeyword{ISM: abundances}

\begin{document}
\maketitle

\section{Introduction}
\label{Sintro}

O/H abundance ratios derived from recombination lines 
of O and H are higher than those derived from the ratio of a  
collisionally excited line (CL) of oxygen to a recombination line 
(RL) of H, this effect is called the abundance discrepancy problem, 
and the ratio of both types of abundances is called the abundance 
discrepancy factor (ADF). This problem also applies to other heavy
elements like C, N, and Ne and is present in \ion{H}{2} regions
as well as in planetary nebulae.

The two main explanations for the large ADF values found in planetary 
nebulae are: a) the presence of large temperature variations, larger 
than those predicted by photoionized models such as CLOUDY \citep{fer13} 
in a chemically homogeneous medium 
\citep[e.g.,][and references therein]{pem06}, and b) the presence of large 
temperature variations due to chemical inhomogeneities 
\citep[e.g.,][and references therein]{liu06}. Both explanations rely on 
the fact that 
RLs are inversely proportional to the temperature and in colder 
regions become brighter relatively to CLs, while CLs increase with temperature 
and become brighter relatively to RLs in hotter regions. In addition 
CLs in the visual range of the spectrum are almost 
non-existent in very cold regions ($T_e \la 2000$ K).

In a chemically homogeneous medium  $I$(O,RL)/$I$(H,RL) is 
proportional to the O/H ratio and is almost independent of the 
electron temperature. Alternatively $I$(O,CL)/$I$(H,RL) depends strongly 
on the electron temperature in such a way that, in the presence of 
temperature variations, the O/H abundances derived from temperature 
determinations based on CLs, assuming constant temperature, 
{are underestimated} giving rise to the presence of
an ADF.

In a chemically inhomogeneous medium CLs are expected to 
originate mainly in regions that are relatively metal-poor, 
temperature-high and density-low, while the RLs are expected to 
originate mainly in regions that are relatively metal-rich, 
temperature-low, and density-high.

There are many heavy element line ratios that can be used to 
determine the physical characteristics of a photoionized region
using CLs (i.e. capable of characterizing the hottest parts of 
photoionized regions). But until recently there were no heavy element 
RL ratios capable of determining physical conditions (i.e. capable 
of characterizing the coldest parts of photoionized regions).
It is difficult, but possible, to determine physical conditions 
using light element RL ratios, this suggests moderate thermal 
inhomogeneities for H and He, but cannot discriminate between 
moderate or large thermal inhomogeneities for heavy elements, 
since, in the presence of chemical inhomogeneities the physical 
determinations made for H or He would not be applicable to O, N, 
Ne, or C.

A decade ago \citet{rui03} \citet{pea05} and \citet{pem05} observationally 
found that the \ion{O}{2} V1 multiplet was not in local thermal 
equilibrium, result that was theoretically confirmed later by \citet{bas06}. The atomic 
physics taking into account the density dependence of the level
populations of the ground states of the  \ion{O}{2} V1 multiplet
has been computed by Storey (\citealt{bas06, liu12, fan13}), 
and is the one used in this paper.

The \ion{O}{2} V1 multiplet consists of 8 lines emitted from atoms in the 4 
energy levels of the \ion{O}{2} 1s$^2$ 2p$^2$ (3P) 3p configuration 
that decay to the 3 energy levels of the 
\ion{O}{2} 1s$^2$ 2p$^2$ (3P) 3s configuration. Since the multiplet 
is optically thin if one can obtain at least one of the 
lines coming from each one of the 4 upper levels it 
is enough to fully characterize the emission from 
the 8 lines of the multiplet. The lines arising from 
each level are: 
from \ion{O}{2} 2p$^2$ (3P) 3p1 1/2 $\lambda\lambda$ 4651 and 4673; 
from \ion{O}{2} 2p$^2$ (3P) 3p1 3/2 $\lambda\lambda$ 4638, 4662, and 4696; 
from \ion{O}{2} 2p$^2$ (3P) 3p1 5/2 $\lambda\lambda$ 4642 and 4676; 
and from \ion{O}{2} 2p$^2$ (3P) 3p1 7/2 $\lambda\lambda$ 4649. 
In LTE the intensity of $\lambda(4649)$ is expected to be 39.7\% of the multiplet \citep{wie96}. 
When densities are lower than $n_e \sim 25000$ cm$^{-3}$ this line becomes 
weaker dropping to 10\% of the multiplet for very low densities. Since 
\ion{O}{2} $\lambda$4649 was expected to be the brightest line 
of the multiplet, and since the lines of the multiplet are all very faint, many studies 
focused in measuring only the \ion{O}{2} $\lambda$4649 line; unfortunately, for most 
photoionized regions,  this results in underestimating the 
oxygen RL abundances by factors of up to 4 if the LTE predictions are adopted \citep{pem05}.

Recent calculations by Storey (presented by 
\citealt[][hereafter PP13]{pem13}, and \citealt{fan13}) 
allow to study the temperature and density dependance of the 
\ion{O}{2} V1 multiplet ratios. These calculations 
indicate that the individual line intensities are density 
and temperature dependent, while the sum is only 
temperature dependent. So far this is the only multiplet 
where the line by line density dependance has been
characterized by both observations and atomic physics calculations.

PP13 studied the relevance of the \ion{O}{2} V1 recombination lines 
for the determination of the physical conditions of \ion{H}{2} regions.
They considered nine galactic and extragalactic H II regions
with high quality line intensity observations. They found that, for 
these objects, the densities derived from CLs (i.e. the densities in the 
hotter zones of the nebulae) are higher than the densities derived from 
RLs (i.e. the densities in the colder zones); this, together with an analysis
of the temperature structure derived from H, He, and O, shows that high 
metallicity inclusions do not contribute much to the observed ADFs in HII 
regions; it also suggests that shockwaves do contribute importantly to the 
ADF presence.

We decided to carry out a similar study to that of PP13 based on the 
best observed planetary nebulae trying to find out if these objects are 
chemically homogeneous or not. And in general to study which 
physical processes are important as sources of thermal inhomogeneities 
in ionized nebulae. 

\section{The sample}\label{Ssample}

We selected 20 PNe from the literature with high quality spectra, both in 
terms of resolution and signal to noise ratio. The objective was to look for 
objects where enough lines of the \ion{O}{2} V1 multiplet were measured 
with high signal to noise ratio to properly characterize the whole multiplet, i.e., 
at least one line from each of the four upper levels that produce this multiplet.

Table~\ref{tab1} lists the studied PNe, the intensity ratios involving 
RLs of the \ion{O}{2} V1 multiplet, and the references from which we took 
the emission line fluxes; F1 stands for the sum of the intensities of the
[\ion{O}{3}] $\lambda$4959 and [\ion{O}{3}] $\lambda$5007 lines of multiplet F1.

{The objects with the best \ion{O}{2} intensity measurements available from 
the literature} produce two selection effects on our sample:
a) Wesson, Liu and collaborators were interested in PNe with high ADF values (8 objects), and b) 
Garc\'ia-Rojas and collaborators were interested in PNe with [WC] central stars 
or with weak emission line central stars (11 objects).

{These two biases imply that the sample presented in this paper may not be representative 
of the total family of PNe, in the sense that the average ADF value for a more general sample
could be smaller than for this sample. To advance further in this subject, it would be important 
to obtain additional observations of similar or higher quality than those used in this paper.}
 
There are other line ratios of \ion{O}{2} recombination lines that can be used to
determine densities and temperatures, see \citet{fan13}. Unfortunately, 
one or both of the lines involved are at least one order of magnitude fainter
than the V1 lines, and therefore the measured line intensities present larger errors. 
In addition at that level of intensity and with the spectral resolutions
available it is not possible to separate the contribution due to other weaker
lines that could be contributing to the measured line intensities.

\begin{table}[!h]\centering
\setlength{\tabnotewidth}{\columnwidth}
\tablecols{6}
\setlength{\tabcolsep}{1.\tabcolsep}
\caption{Sample of PNe and line intensity ratios.} \label{tab1}
\begin{tabular}{lr@{$\pm$}lr@{$\pm$}lc}
\toprule
\multicolumn{1}{l}{Object} & \multicolumn{2}{c}{\TabV{$\frac{\displaystyle I({\rm V}1)\tabnotemark{a}}{\displaystyle I({\rm F}1)}$}}  
& \multicolumn{2}{c}{\TabV{$\frac{\displaystyle I(4649)\tabnotemark{b}}{\displaystyle I(4639+51+62)}$}} & \multicolumn{1}{c}{Ref.\tabnotemark{c}} \\
\midrule
Cn~1-5     & 97  &  12    &   115  & 29 & (1)     \\
He~2-86   & 122  &  4    & 122  & 6 &  (1)    \\
Hu~1-1     &  28  &  11  &  76  & 37 &  (2)   \\
Hu~2-1     &  116  &  9   & 136 & 16  &  (2)    \\
M~1-25     &  158  &  15   & 82  & 17 & (1)   \\
M~1-30     & 454  &  20  &  92 & 9 & (1)  \\
M~1-61     &   80   &  4   & 120 & 19  &  (1)   \\
M~3-15     &   149   &  22   & 87  & 28 &  (1)    \\
NGC~2867   &  51  &  2    & 41 & 5  &  (3)  \\
NGC~5189   &  39  &  4   & 61  & 13 &  (1)   \\
NGC~5307   &  38  &  4   & 73  & 15 &  (4) \\
NGC~6153   &  346  &  10   & 106 & 5  & (5) \\
NGC~6803   &  89  &  5   & 116 & 12  &  (2)   \\
NGC~6879   &  56  &  12   & 30 & 14  &  (2)   \\
NGC~6891   &  68  &  5   & 68 & 14  & (2)   \\
NGC~7009   &  133  &   10  & 92 & 11  & (6) \\
NGC~7026   &  110  &  5   &  106 & 11  & (2)   \\
PB~8       &  365 &  16    & 71 & 8  &  (3) \\
PC~14     &  82  &  6    & 88 & 12  &  (1)   \\
Pe~1-1    &  57  &  4   &\hspace{0.7cm}   127 & 30  \hspace{0,7cm}  & (1)   \\
\bottomrule
\tabnotetext{a}{In units of $10^{-5}$.}
\tabnotetext{b}{In units of $10^{-2}$.}
\tabnotetext{c}{References--- (1) \citet{gar12}, (2) \citet{wes05}, (3) \citet{gar09}, (4) \citet{rui03}, 
(5)  {\citet{liu00}}, (6) \citet{fan11}.}
\end{tabular}
\end{table}

\section{Physical conditions}

\subsection{Temperature determinations based on \ion{O}{2} recombination lines}\label{Scz-temp}

Based on the atomic data for case B by \citet{sto94,sto00, sto14} we find that
\begin{eqnarray}
\label{Ere-V1/4959}
\nonumber  {I({\rm V1}) \over I({\rm F1})} =\hspace{4.5cm}\\
 \hspace{0cm}1.772 \times 10^{-5} \left(\frac{T_{\rm e}}{10000 \,\rm K}\right)^{-0.40}\exp ({29170 \,\rm K}/T_{\rm e}),
\end{eqnarray}
We will call \temp(V1/F1) the temperature derived from equation~(\ref{Ere-V1/4959}). 
These temperatures are presented in Table~\ref{Tt-oii}.

\begin{table}[!h]\centering
\setlength{\tabnotewidth}{\columnwidth}
\tablecols{5}
\setlength{\tabcolsep}{2.\tabcolsep}
\caption{Physical conditions of the cold zones.} \label{Tt-oii}
\begin{tabular}{lr@{}lr@{}l}
\toprule
\multicolumn{1}{c}{Object} & \multicolumn{2}{c}{\temp(V1/F1)} & \multicolumn{2}{c}{\dens(\ion{O}{ii})}\\
\midrule
Cn~1-5        & 7500 & $\pm195$   & 8880 & $^{+\infty}_{-5130}$  \\
He~2-86        & 7120 & $\pm55$     & 13300 & $^{+6730}_{-3590}$  \\ 
Hu~1-1         & 9790 & $\pm740$  & 2480 & $^{+6240}_{-1350}$  \\ 
Hu~2-1         & 7270 & $\pm125$  & 49790 & $^{+\infty}_{-37040}$ \\
M~1-25         & 6740 & $\pm125$  & 2500 & $^{+2110}_{-890}$ \\ 
M~1-30         & 5500 & $\pm40$    & 3230 & $^{+1320}_{-820}$  \\
M~1-61         & 7860 & $\pm100$  & 11760 & $^{+154500}_{-6120}$ \\
M~3-15         & 6810 & $\pm200$  & 3030 & $^{+5640}_{-1490}$ \\ 
NGC~2867    & 8850 & $\pm120$ &  710 & $^{+160}_{-130}$  \\
NGC~5189    & 9510 & $\pm260$ & 1530 & $^{+770}_{-470}$ \\
NGC~5307    & 9600 & $\pm250$ &  2210 & $^{+1280}_{-690}$ \\  
NGC~6153    & 5800 & $\pm30$   & 5540&$^{+1160}_{-940}$  \\ 
NGC~6803    & 7540& $\pm70$    & 9360  & $^{+10240}_{-3420}$ \\
NGC~6879    & 7930 & $\pm290$ & 400 & $^{+380}_{-240}$  \\
NGC~6891    & 8460 & $\pm165$ & 1800 & $^{+940}_{-550}$ \\
NGC~7009    & 6990 &$\pm115$  & 3590&$^{+1660}_ {-1020}$ \\ 
NGC~7026    & 7160 & $\pm70$   & 5890 &$^{+4160}_ {-1840}$  \\
PB~8              & 5710 & $\pm45$  & 1630 & $^{+450}_{-340}$  \\
PC~14            & 7810 & $\pm125$ & 3290 & $^{+1680}_{-970}$  \\
Pe~1-1           & 8550 & $\pm135$ & 17400 & $^{+\infty}_{-11920}$ \\
\bottomrule
\end{tabular}
\end{table}

Equation~(\ref{Ere-V1/4959}) is density independent because  the sum  of the intensities of the 
eight lines of multiplet V1 is density independent. Moreover, equation~(\ref{Ere-V1/4959}) is 
different to equation (6) in PP13 for two reasons: a) we include the [\ion{O}{3}] $\lambda$5007 
line in the calculations, and b) the equation in PP13 had a misprint, \temp$^{-0.415}$ 
should have been \temp$^{-0.415}$/10000 K.

The nebular lines have a much weaker temperature dependence than the auroral lines; 
the V1 multiplet temperature dependence is approximately equally strong to that of the 
nebular lines, but biased towards colder temperatures. Therefore, in the presence 
of temperature inhomogeneities, \temp(V1/F1) will represent better the cold zones 
of the nebulae than the normal auroral versus nebular temperatures, so we will use 
\temp(V1/F1) to represent them.

\subsection{Density determinations based on \ion{O}{2} recombination lines}\label{Scz-dens}

It is possible to determine densities using only lines from the \ion{O}{2} V1 multiplet \citep{rui03}.
From Figure 3 of PP13, that is based on the unpublished computations by Storey
(\citealt{bas06,liu12,fan13}), and the $I$(4649)/$I$(4639+51+62) ratio presented in Table~\ref{tab1}, 
we have computed the $n_e$(\ion{O}{2}) values presented in Table \ref{Tt-oii}. 
For those objects where this ratio is equal or higher than 1.15, and the error is greater than 10\% 
(Cn1-5, Hu~2-1, M~1-61, and Pe~1-1), {the derived densities are very uncertain, with no real 
restriction in the upper limit.} 

For all the other objects, this determination requires a temperature to uniquely derive the density. 
Since RLs are brighter in the cold zones of photoionized regions, a colder temperature than the 
one derived from nebular versus auroral lines is required. For the chemically homogeneous 
model the correct temperature to use can be approximated by \temp(V1/F1). In the 
presence of high-metallicity, dense clumps a lower temperature should be used; with lower 
temperatures the derived densities will also be lower (see Figure 3 of PP13) by approximately 
\dens $\propto$ \temp$^{2/3}$. We will discuss the implications of this result in \S4.

\subsection{Temperature and densities based on collisionally excited lines}\label{Shz-phys}

We have computed \temp([\ion{O}{3}]) and \dens([\ion{Cl}{3}]) as representative values 
of the temperature and density in the hot regions of the O$^{++}$ zone. They were 
calculated from the [\ion{O}{3}] ($\lambda$4959+$\lambda$5007)/$\lambda$4363 
and  [\ion{Cl}{3}] $\lambda\lambda$5517, 5537 diagnostic ratios respectively. 

For the four PNe where [\ion{Cl}{3}] lines were not available (Hu~1-1, Hu~2-1, NGC~6879, 
and NGC~6891), we adopted the average value of the densities obtained with other three 
diagnostic ratios: [\ion{O}{2}] $\lambda$3727/$\lambda$3729, [\ion{S}{2}] $\lambda$6716/$\lambda$6731, 
and [\ion{Ar}{4}] $\lambda$4711/$\lambda$4740. This assumption is reasonable since 
in the other 16 PNe where all the diagnostic ratios are available, the agreement between 
the average density and the one obtained from chlorine lines is better than $\sim$30 \%.

All the calculations have been performed with the software PyNeb \citep{lur12} and the uncertainties 
associated with the physical conditions have been computed through Montecarlo simulations. {We 
adopted the transitions probabilities and the collision strengths from \citet{men82b} and \citet{ram97} 
for Ar$^{+3}$, \citet{men82a} and \citet{but89} for Cl$^{++}$, \citet{sto00} and \citet{sto14} for O$^{++}$, 
\citet{wie96} and \citet{kis09} for O$^{+}$, and \citet{zei82} and \citet{tay10} for S$^{+}$.}
 
Table~\ref{Tt-oiii} shows the final temperatures and densities derived for each nebula. The differences 
between our values and those provided in the papers listed in Table~\ref{tab1} are about 3\% 
for the \temp's and about 15\% for the \dens's, with a few exceptions showing higher differences. 
The differences are caused by the different sets of atomic data used in the calculations. 

\begin{table*}[!t]\centering
\setlength{\tabnotewidth}{\columnwidth}
\tablecols{7}
\setlength{\tabcolsep}{2.\tabcolsep}
\caption{Physical conditions of the hot zones and pressure ratios between 
the hot and cold zones.} \label{Tt-oiii}
\begin{tabular}{lr@{}lr@{}lr@{}l}
\toprule
\multicolumn{1}{c}{Object} & \multicolumn{2}{c}{\temp([\ion{O}{iii}])} & \multicolumn{2}{c}{\dens([\ion{Cl}{iii}])} 
& \multicolumn{2}{c}{$\log P({\rm hz})/P({\rm cz})$}\\
\midrule
Cn~1-5         & 8850&$\pm165$   &  3400&$_{-700}^{+800}$  & $-0.36$ & $_{-\infty}^{+0.38}$  \\
He~2-86       & 8500&$\pm150$   &  17350&$_{-3100}^{+4300}$ & $0.18$ & $_{-0.23}^{+0.15}$  \\
Hu~1-1         & 11870&$\pm270$  &  1500&$\pm300$  & $-0.14$  & $_{-0.56}^{+0.35}$ \\
Hu~2-1         & 10890&$\pm260$  &  10600&$_{-4000}^{+9300}$ & $-0.50$  & $_{-\infty}^{+0.61}$ \\
M~1-25         & 7840&$\pm145$   &  11650&$_{-2100}^{+2200}$ & $0.75$  & $_{-0.28}^{+0.21}$ \\
M~1-30         & 6650&$\pm150$   &  6200&$_{-1000}^{+1300}$ & $0.37$  & $_{-0.17}^{+0.15}$ \\
M~1-61         & 9260&$\pm200$   &  16000&$_{-3200}^{+4300}$ & $0.18$  & $_{-1.16}^{+0.33}$  \\
M~3-15         & 8400&$\pm225$   &  7700&$_{-1400}^{+1800}$ & $0.49$  & $_{-0.47}^{+0.30}$ \\
NGC~2867    & 11900&$\pm285$  &  3800&$_{-800}^{+900}$ & $0.87$ & $\pm0.13$   \\
NGC~5189    & 11670&$\pm280$  &  1150&$_{-400}^{+500}$ & $-0.04$ & $_{-0.27}^{+0.22}$  \\
NGC~5307    & 11770&$\pm90$    &  1600&$_{-850}^{+950}$ & $0.13$ & $_{-0.21}^{+0.18}$   \\ 
NGC~6153    &  9120&$\pm155$    &  3550&$_{-600}^{+700}$ & $-0.003$ & $_{-0.13}^{+0.11}$  \\
NGC~6803    & 9650&$\pm180$   &  8700&$_{-1200}^{+1600}$ & $0.08$ & $_{-0.33}^{+0.21}$ \\
NGC~6879    & 12010&$\pm270$  &  4450&$_{-700}^{+1100}$ & 1.23 & $_{-0.31}^{+0.40}$ \\
NGC~6891    & 9180&$\pm160$   &  1850&$\pm400$\ & 0.05 & $_{-0.21}^{+0.18}$ \\
NGC~7009    & 9900&$\pm180$  & 3400&$\pm200$  & 0.12  & $_{-0.17}^{+0.15}$ \\
NGC~7026    & 9390&$\pm165$   &  8800&$_{-1300}^{+1500}$ & 0.29 & $_{-0.24}^{+0.18}$  \\
PB~8             & 6940&$\pm110$   &  1750&$_{-900}^{+1400}$ & 0.16 & $_{-0.56}^{+0.21}$ \\
PC~14           & 9370&$\pm170$   &  2900&$_{-550}^{+800}$ & 0.04 & $_{-0.21}^{+0.17}$ \\
Pe~1-1          & 10080&$\pm235$  &  24000&$_{-5000}^{+7550}$ & 0.21 & $_{-\infty}^{+0.51}$ \\
\bottomrule
\end{tabular}
\end{table*}

\section{Densities and ADF\MakeLowercase{s}}

\begin{figure}[!h]
\begin{centering}
\includegraphics[width = \hsize, trim = 20 20 55 10, clip = yes]{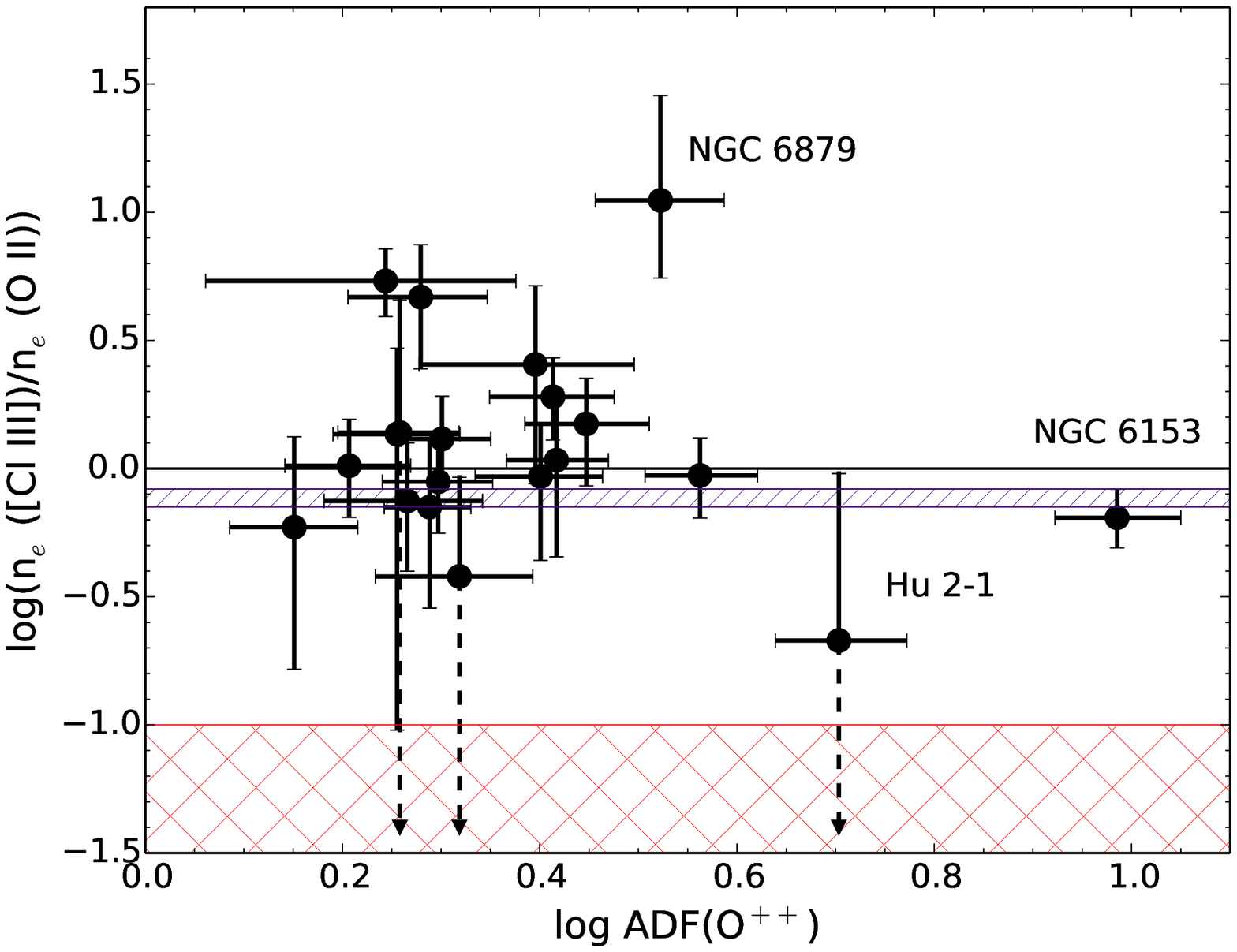}
\caption{Ratio of \dens([\ion{Cl}{3}]) to \dens(\ion{O}{2}) values as a function of ADF(O$^{++}$). 
The solid line shows where the values of \dens(\ion{O}{2}) and \dens([\ion{Cl}{3}]) are equal. 
The small dashed area represents a chemically homogeneous region of constant pressure. 
The crossed area at the bottom represents models with constant pressure in a 
chemically inhomogeneous medium where the metal poor regions are at least 10 times 
hotter than the metal rich regions.\label{fig1}}
\end{centering}
\end{figure}

Figure~\ref{fig1} shows the ratio of \dens(\ion{O}{2}) to \dens([\ion{Cl}{3}]) values 
as a function of the ADF(O$^{++}$), given by
\begin{equation}
{\rm ADF}({\rm O}^{++}) = \frac{({\rm O}^{++}/{\rm H}^+)_{\rm RLs}}{({\rm O}^{++}/{\rm H}^+)_{\rm CLs}}.
\end{equation}
The values of O$^{++}$/H$^+$ from RLs were directly taken from the V1 multiplet abundances 
presented in the papers listed in Table~\ref{tab1}. 
In those PNe where no uncertainties are provided by the authors we assume a one sigma error of $\pm$0.05 dex.
The O$^{++}$ abundances from CLs were computed with the [\ion{O}{3}] $\lambda\lambda$4959, 5007/H$\beta$ 
intensity ratios and the physical conditions of Table~\ref{Tt-oiii}. 
They are listed in Table~\ref{Tadf} together with the ADF(O$^{++}$) values.

\begin{table}[!t]\centering
\setlength{\tabnotewidth}{\columnwidth}
\tablecols{7}
\setlength{\tabcolsep}{1.\tabcolsep}
\caption{O$^{++}$ abundances and ADF} \label{Tadf}
\begin{tabular}{lr@{$\pm$}lr@{}lr@{}l}
\toprule
\multicolumn{1}{c}{Object} & \multicolumn{2}{c}{\{O$^{++}$\}\tabnotemark{a}$_{\rm RLs}$} & \multicolumn{2}{c}{\{O$^{++}$\}$_{\rm CLs}$} & \multicolumn{2}{c}{ADF(O$^{++}$)}\\
\midrule
Cn~1-5         & 8.99 & 0.06 & 8.67&$\pm0.04$ & 2.08&$^{+0.39}_{-0.37}$ \\
He~2-86       & 9.02 & 0.02 & 8.72&$\pm0.04$ & 2.00&$^{+0.24}_{-0.20}$   \\ 
Hu~1-1         & 8.57 & 0.05 & 8.42&$\pm0.04$ & 1.42&$^{+0.23}_{-0.20}$  \\ 
Hu~2-1         & 8.73 & 0.05 & 8.03&$\pm0.04$ & 5.05&$^{+0.87}_{-0.69}$ \\
M~1-25         & 8.89 & 0.05 & 8.61&$\pm0.04$ & 1.90&$^{+0.32}_{-0.30}$   \\ 
M~1-30         & 8.92 & 0.03 & 8.51&$\pm0.05$ & 2.59&$^{+0.40}_{-0.36}$   \\
M~1-61         & 8.88 & 0.04 & 8.62&$\pm0.04$ &  1.80&$^{+0.28}_{-0.25}$   \\
M~3-15         & 9.17 & 0.09 & 8.77&$\pm0.05$ & 2.49&$^{+0.64}_{-0.59}$\\ 
NGC~2867   & 8.65 & 0.13 & 8.41&$\pm0.04$ & 1.75&$^{+0.62}_{-0.60}$   \\
NGC~5189   & 8.66 & 0.06 & 8.40&$\pm0.04$ & 1.84&$^{+0.35}_{-0.33}$   \\
NGC~5307   & 8.77 & 0.04 & 8.48&$\pm0.01$ & 1.94&$\pm0.20$  \\  
NGC~6153   & 9.61 & 0.05 & 8.62&$\pm0.04$ & 9.67&$^{+1.55}_{-1.31}$  \\ 
NGC~6803   & 9.07 & 0.05 & 8.67&$\pm0.04$ & 2.52&$^{+0.39}_{-0.36}$  \\
NGC~6879   & 8.82 & 0.05 & 8.30&$\pm0.04$ & 3.33&$^{+0.53}_{-0.47}$  \\
NGC~6891   & 8.82 & 0.05 & 8.61&$\pm0.04$ & 1.61&$^{+0.25}_{-0.22}$  \\
NGC~7009   & 9.18  & 0.05 & 8.62&$\pm0.02$ & 3.65&$^{+0.52}_{-0.44}$  \\ 
NGC~7026   & 9.08 & 0.05 & 8.64&$\pm0.04$ & 2.80&$^{+0.44}_{-0.37}$ \\
PB~8            & 9.15 & 0.03 & 8.73&$\pm0.04$ & 2.61&$^{+0.34}_{-0.29}$ \\
PC~14          & 9.01 & 0.04 & 8.72&$\pm0.04$ & 1.98&$^{+0.27}_{-0.24}$ \\
Pe~1-1         & 8.79 & 0.04 & 8.53&$\pm0.04$ & 1.81&$^{+0.27}_{-0.24}$\\
\bottomrule
\tabnotetext{a}{From the papers listed in Table~\ref{tab1}.}
\end{tabular}
\end{table}

The ADF is present in all the objects of our sample. The ADF 
range goes from 1.42 (Hu~1-1) to 9.67 (NGC~6153), being the average value 2.74. 
If we eliminate the two objects with the highest ADF value, the
range of the other 18 goes from 1.42 to 3.65 with an average value of
2.23. The average value {would be probably} smaller for a typical sample of PNe,
because part of our sample is biased towards objects with high ADF values.

Two sets of models have been proposed to explain the ADF values: 
chemically inhomogeneous models and chemically homogeneous ones.

Chemically inhomogeneous models consist of pockets of high density
and low temperature embedded in a medium of lower density and higher
temperature. In these models most of the mass is located in the medium
of lower density and higher temperature and consequently the
proper abundances for heavy  elements are those obtained from the
intensity ratios of forbidden lines to hydrogen recombination lines.

A chemically homogeneous model requires the consideration of additional 
physical processes, to those given by direct ionization from the central star, 
to explain the observed ADF values. Four of these processes are: shocks, 
magnetic reconnection, shadowed ionization, and a receding ionization front.

In the presence of large temperature variations the ratio of a collisionally 
excited line to a recombination line  depends
strongly on the temperature. In the presence of temperature inhomogeneities
when the temperature is derived from the ratio of two collisionally
excited lines the abundances of the heavy elements derived from the ratio
of a collisionally excited line to a hydrogen line
are lower limits to the real abundance ratio \citep{pem67, pem69}.

On the other hand the recombination lines of the heavy elements and
of hydrogen depend weakly on the electron temperature, they are
inversely proportional to the temperature and to a very good approximation
the intensity ratios of two recombination lines are almost independent
of the average temperature as well as of the presence of temperature 
inhomogeneities. Therefore in models with temperature inhomogeneities
the proper abundances are those provided by recombination
lines of the heavy elements to those of hydrogen.

From Tables~\ref{Tt-oii} and \ref{Tt-oiii} and Figure~\ref{fig1} it follows that six objects 
have \dens([\ion{Cl}{3}]) higher than the \dens(\ion{O}{2}) values and 11 objects have \dens([\ion{Cl}{3}]) 
values that are consistent with the \dens(\ion{O}{2}) values; the three other objects, Cn~1-5, Hu~2-1, and 
NGC~6153, are not compatible with density equilibrium but are {consistent} with pressure equilibrium. 
However, it can not been rule out at one sigma level that four objects (Cn~1-5, Hu~2-1, 
M~1-61, Pe~1-1) may have significantly low \dens([\ion{Cl}{3}])/\dens(\ion{O}{2}) ratios. 

The minimum temperature that we are adopting for the \dens(\ion{O}{2}) determinations is 5500 K. 
In the chemical inhomogeneous model the temperature of the cold zones is expected to be much 
lower than this. The chemically inhomogeneous models are expected to be able to reproduce the 
observed V1/F1 ratio by having a much lower temperature in the cold zone. 
If the temperatures were lower than those adopted 
in this paper, the measured \dens(\ion{O}{2}) value would decrease even further. 
This is not compatible with the inhomogeneous models that usually require the 
densities of the cold zones to be at least one order of magnitude higher than those of 
the hot zones. 

The other interesting point that follows from Figure~\ref{fig1} is that there is no correlation between 
\dens([\ion{Cl}{3}])/\dens(\ion{O}{2}) and the ADF(O$^{++}$) value. For chemically inhomogeneous 
PNe it is expected that the higher the ADF(O$^{++}$) value the lower the  \dens([\ion{Cl}{3}])/\dens(\ion{O}{2}) 
ratio.

\section{Temperature structure}

The {\temp([\ion{O}{iii}])}/{\temp(V1/F1)} ratio varies in the 1.09 to 1.58 range with
an average value of 1.28, the most extreme object is NGC 6153.

We decided to follow the formalism introduced by \citet{pem67} 
to determine the basic parameters of the temperature structure, 
$T_0(\rm{O^{++}})$ and $t^2(\rm{O^{++}})$, where 
\begin{equation}
\label{Edef-T0}
T_0({\rm O^{++}})=
\frac{\int T_e n_e n({\rm O^{++}})dV}
{\int n_e n({\rm O^{++}})dV},  
\end{equation}
and
\begin{equation}
\label{Edef-t2}
t^2({\rm O^{++}})=
\frac{\int (T_e - T_0({\rm O^{++}}))^2 n_e n({\rm O^{++}})dV}
{T_0({\rm O^{++}})^2 \int n_e n({\rm O^{++}})dV}.  
\end{equation}

To study $T_0$(O$^{++}$) and $t^2$(O$^{++}$) we need an analytical expression for \temp([\ion{O}{3}]).
Based on the atomic data by \citet{sto00, sto14} we find that
\begin{equation}
\label{Ere-4363/4959}
{I(4363) \over I(F1)} =
0.1308\exp ({-32965 \,\rm K}/T_{\rm e}).
\end{equation}
From equations (\ref{Ere-V1/4959}), (\ref{Edef-T0}), (\ref{Edef-t2}), and (\ref{Ere-4363/4959}), 
and the temperature dependence of the emissivities
$\varepsilon_{\rm V1}$, $\varepsilon_{4959}$, and 
$\varepsilon_{4363}$ (see PP13), we can write $T_e$(V1/F1) and $T_e([\ion{O}{3}])$ as a
function of $T_0$ and $t^2$ as follows:
\begin{eqnarray}
\label{ET0-OIII}
\nonumber T_e([\ion{O}{3}]) = T_0({\rm O^{++}})\hspace{3.5cm}\\
\times\left[ 1 + \left( \frac{91305 \,\rm K}{T_0({\rm O^{++}})} - 2.74\right)
\frac{t^2({\rm O^{++}})}{2} \right],
\end{eqnarray}
and
\begin{eqnarray}
\label{ET0-OII}
\nonumber T_e({\rm V1}/{\rm F1}) = T_0({\rm O^{++}}) \hspace{5cm}\\
\times\left[ 1+\left( \frac{29170 \,\rm K}{T_0({\rm O^{++}})} - 3.14 + 
\frac{0.40}{\frac{29170 \,\rm K}{T_0({\rm O^{++}})} +0.40} \right) {t^2({\rm O^{++}}) \over 2} \right].
\end{eqnarray}
Using equations (\ref{ET0-OIII}) and (\ref{ET0-OII}), along with the temperatures presented in 
Tables~\ref{Tt-oii} and \ref{Tt-oiii}, we have derived the $T_0$(O$^{++}$) and $t^2$(O$^{++}$) 
values presented in Table \ref{Tt2}.

\begin{table*}[!t]\centering
\setlength{\tabnotewidth}{\columnwidth}
\tablecols{9}
\setlength{\tabcolsep}{2.\tabcolsep}
\caption{Other temperatures and {\MakeLowercase {\it t}}$^2$ values} \label{Tt2}
\begin{tabular}{lr@{}lr@{}lr@{}lr@{}l}
\toprule
\multicolumn{1}{c}{Object} & \multicolumn{2}{c}{$T_0$(O$^{++}$)} & \multicolumn{2}{c}{$T$(Bac/Pac)\tabnotemark{a}} 
& \multicolumn{2}{c}{$t^2$(O$^{++}$)} & \multicolumn{2}{c}{$t^2$(\ion{He}{1}/CL)\tabnotemark{a}}\\
\midrule
Cn~1-5         & 7305&$\pm235$ & $\ldots$ &  & 0.043&$\pm0.008$ & 0.042&$\pm0.014$  \\
He~2-86       & 6905 & $\pm70$ & 7560 & $^{+3050}_{-2060}$ & 0.044&$\pm0.005$ & 0.036&$\pm0.014$ \\ 
Hu~1-1         & 9700 & $\pm780$ & 8350 &  &  0.067&$\pm0.024$ & $\ldots$&  \\ 
Hu~2-1         & 6735 & $\pm180$ & 8960 &  &  0.113&$\pm0.009$ & $\ldots$&  \\
M~1-25         & 6545 & $\pm155$ & 7750 & $^{+3100}_{-2050}$ & 0.035&$\pm0.006$ & 0.035&$\pm0.019$  \\ 
M~1-30         & 5240 & $\pm60$ & 5800 & $^{+2150}_{-1300}$ & 0.037&$\pm0.005$ & 0.013&$^{+0.018}_{-0.013}$ \\
M~1-61         & 7675 & $\pm120$ & 9800 & $^{+4150}_{-2700}$ & 0.044&$\pm0.006$ & 0.034&$\pm0.007$  \\
M~3-15         & 6560 & $\pm255$ & 9800 & $^{+4150}_{-2800}$ & 0.049&$\pm0.010$ & 0.078&$\pm0.030$ \\ 
NGC~2867    & 8625 & $\pm150$ & 8950 & $^{+2900}_{-1900}$ &  0.096& $\pm0.007$ & 0.046&$\pm0.029$ \\
NGC~5189    & 9390 & $\pm295$ & 9200 & $^{+3600}_{-2300}$  & 0.069&$\pm0.011$ & 0.054&$\pm0.008$  \\
NGC~5307    & 9490 & $\pm280$ & 10700 & $\pm2000$ & 0.069&$\pm0.008$ & 0.031&$\pm0.014$ \\  
NGC~6153    & 5050 & $\pm60$ & 6080&  &  0.105&$\pm0.005$ & $\ldots$& \\ 
NGC~6803    & 7250 & $\pm90$ & 7320 &  &  0.067&$\pm0.006$ & $\ldots$& \\
NGC~6879    & 7440 & $\pm400$ & 8500 & & 0.128&$\pm0.012$ & $\ldots$& \\
NGC~6891    & 8350 & $\pm185$ & 5930 &  &  0.024&$\pm0.008$ & $\ldots$& \\
NGC~7009    & 6515 & $\pm155$ & 6490 &  &  0.092&$\pm0.007$ & $\ldots$&  \\ 
NGC~7026    & 6820 & $\pm95$ & 7440 &  &  0.070&$\pm0.005$ & $\ldots$&  \\
PB~8              & 5455 & $\pm60$ & 5100 & $^{+1300}_{-900}$ &  0.039&$\pm0.004$ & 0.008&$^{+0.020}_{-0.028}$  \\
PC~14            & 7600 & $\pm150$ & 8500 & $^{+3250}_{-2050}$ & 0.050&$\pm0.007$ & 0.040&$\pm0.004$ \\
Pe~1-1           & 8385 & $\pm155$ & 10300 & $^{+4500}_{-2950}$ & 0.051&$\pm0.009$ & 0.044&$\pm0.007$ \\
\bottomrule
\tabnotetext{a}{From the papers listed in Table~\ref{tab1}.}
\end{tabular}
\end{table*}

\begin{figure}[!h]
\begin{centering}
\includegraphics[width = \hsize, trim = 25 20 40 10, clip = yes]{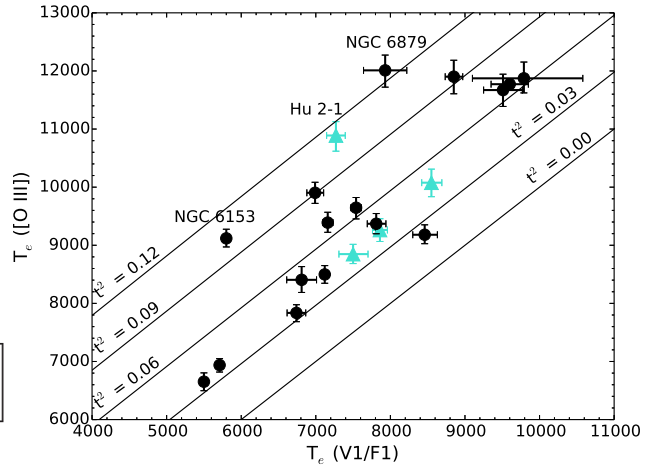}
\caption{Comparison between \temp([\ion{O}{3}]) and \temp(V1/F1). The solid 
lines are sequences of equal $t^2$ values obtained from equations~(\ref{ET0-OIII}) 
and (\ref{ET0-OII}). The four triangles are the objects where we cannot discard the 
presence of chemical inhomogeneities (see Figure~\ref{fig1}). \label{fig2}}
\end{centering}
\end{figure}

Figure~\ref{fig2} shows the comparison between the temperatures of the hot zones, \temp([\ion{O}{3}]), 
and the temperatures of the cold zones, \temp(V1/F1). We overplot in the figure five sequences of equal 
$t^2$. 

This figure is based on the assumption that the chemical composition of each PNe of our 
sample is homogeneous. From this figure we find $t^2$(O$^{++}$) values in the 0.024 to 0.128 range, 
range that is similar to the one obtained from galactic and extragalactic H II regions by \citet{pem12} 
that amounts to 0.019 -- 0.120. The average $t^2$(O$^{++}$) value of our sample is 0.0645, a value 
somewhat higher than that obtained from galactic and extragalactic H II regions by \citet{pem12} that 
amounts to 0.044. 

The values of $t^2$(O$^{++}$) calculated by us and the available $t^2$(He I/CL) values from the literature 
are listed in Table~\ref{Tt2}.

For the 12 PNe of our sample with $t^2$(He I/CL) values in the literature (those observed by 
\citealt{gar09,gar12} and \citealt{rui03}), the average $t^2$(He I/CL) amounts
to 0.042, in very good agreement with the average $t^2$(O$^{++}$) that amounts to 0.045. 
This result supports the idea that O and He are well mixed in these objects.

\section{The role of pressure}

Physically, it is more meaningful to study the pressure ratio than the density ratio 
between the hot and cold zones in nebulae, therefore we will proceed to determine 
the pressure ratio.

As in PP13, the electron pressures were derived through the ideal gas equation, 
$P_{\rm e}$ = \dens$k$\temp. The pressures of the cold zones, $P_{\rm e}$(cz), were obtained with 
\temp(V1/F1) and \dens(\ion{O}{2}), whereas the pressures of the hot zones, $P_{\rm e}$(hz), 
were obtained with \temp([\ion{O}{iii}]) and \dens([\ion{Cl}{iii}]). The last column in 
Table~\ref{Tt-oiii} lists the ratio between both pressures for each nebula. 

In the model with chemical inhomogeneities the predicted high density
pockets of low temperature are expected to reach pressure equilibrium
with the low density medium of high temperature. 
We find that in most of the objects $P_{\rm e}$(hz) is higher than $P_{\rm e}$(cz). This
excess has two implications: a) that probably it is due to the presence of
shock waves, where we expect an increase of the temperature
and the density, and b) that the low densities found in the cold zones would require 
a very large fraction of the O$^{++}$ in the cold zones to reproduce the observed 
recombination line intensities, contrary to the predictions of the chemical inhomogeneous 
models. 

\begin{figure}
\begin{centering}
\includegraphics[width = \hsize, trim = 20 0 40 40, clip = yes]{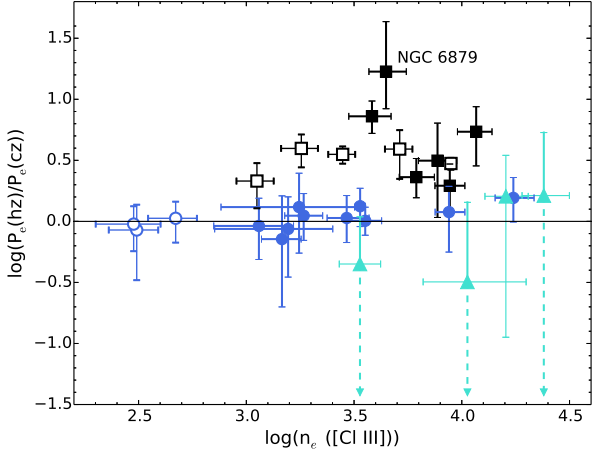}
\caption{Ratio between the pressure in the hot zones, $P_{\rm e}$(hz), and the pressure in the cold zones, $P_{\rm e}$(cz), as a function 
of \dens([\ion{Cl}{3}]). The filled symbols represent PNe and the open ones H~II regions. The squares represent the objects with 
pressure ratios strictly above zero. The circles are those objects with pressure ratios compatible with zero within errors. 
The four triangles are the objects where we cannot discard the presence of chemical inhomogeneities. 
The solid line shows where the pressures of the cold and hot zones are equal. \label{fig3}}
\end{centering}
\end{figure}

\begin{figure}
\begin{centering}
\includegraphics[width = \hsize, trim = 20 15 40 5, clip = yes]{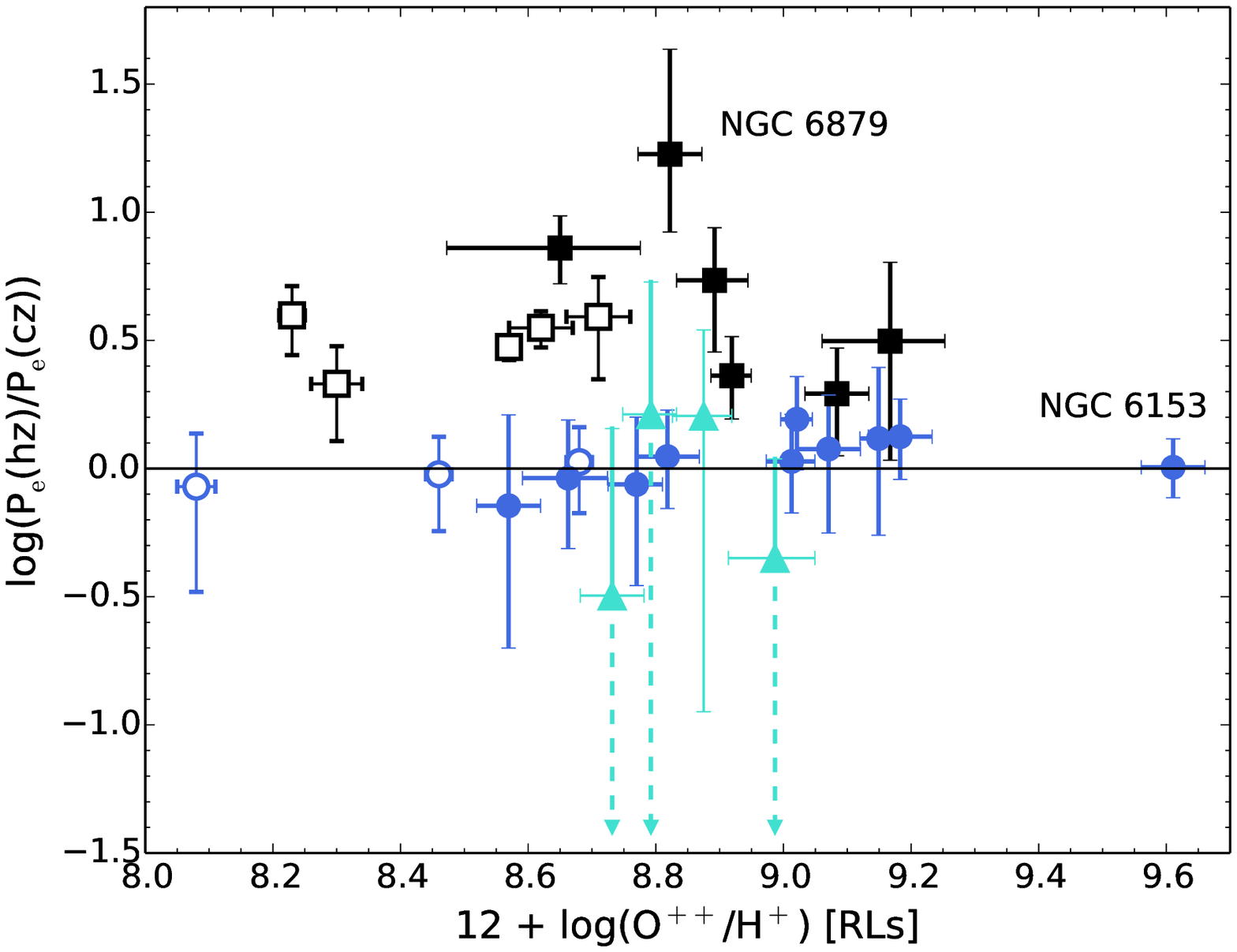}
\caption{Ratio between the pressure in the hot zones, $P_{\rm e}$(hz), and the pressure in the cold zones, $P_{\rm e}$(cz),  
as a function of O$^{++}$/H$^+$ derived from RLs. The symbols and the solid line are the same as in Fig.~\ref{fig3}. \label{fig4}}
\end{centering}
\end{figure}

\begin{figure}
\begin{centering}
\includegraphics[width = \hsize, trim = 20 0 40 5, clip = yes]{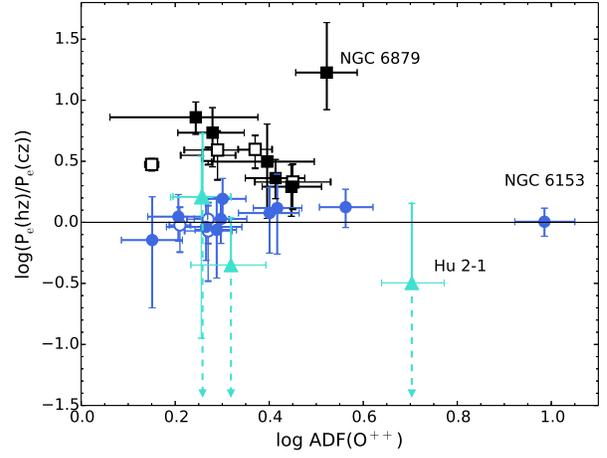}
\caption{Ratio between the pressure in the hot zones, $P_{\rm e}$(hz), and the pressure in the cold zones, $P_{\rm e}$(cz), 
as a function of ADF(O$^{++}$). The symbols and the solid line are the same as in Fig.~\ref{fig3}. \label{fig5}}
\end{centering}
\end{figure}

Figures~\ref{fig3}--\ref{fig5} show the ratio of $P_{\rm e}$(hz) to $P_{\rm e}$(cz) values as a function of 
\dens([\ion{Cl}{3}]), ADF(O$^{++}$), and O$^{++}$/H$^+$ derived from RLs. For comparison we 
include the eight Galactic H~II regions studied in PP13 (\citealt{est04, gar04, gar05, gar06, gar07}).

From Figure~\ref{fig3} we see that nebulae with \dens([\ion{Cl}{3}]) $\gtrsim3200$ cm$^{-3}$
show the largest departures from pressure equilibrium, while the objects with a relatively low 
density seem to be closer to pressure equilibrium than those with higher densities. A similar 
result was found for H~II regions by PP13. As in H~II regions, this result can be related to age 
of the nebulae.

The three PNe with the highest $t^2$ values are NGC~6879, Hu~2-1, and NGC~6153. 
The first one is the PN with the highest pressure ratio, in Hu~2-1 we can not rule out the 
presence of high density and low temperature regions, and NGC~6153 shows pressure equilibrium. 
This indicates that there is no obvious correlation between temperature fluctuations and pressure 
equilibrium. This is also shown in Figure~\ref{fig5}, where the pressure ratios and the ADF show 
no clear relation. In particular, the PN with the highest ADF, NGC~6153, presents similar pressures 
in the the hot and the cold zones. 

The studied PNe and H~II regions cover a wide range in $12+\log$(O$^{++}$/H$^+$), 
from $\sim$8.08 (S311) to 9.61 (NGC~6153). There is no corelation between the pressure ratios 
and the O$^{++}$ abundances.

\begin{figure}
\begin{centering}
\includegraphics[width = \hsize, trim = 20 20 40 20, clip = yes]{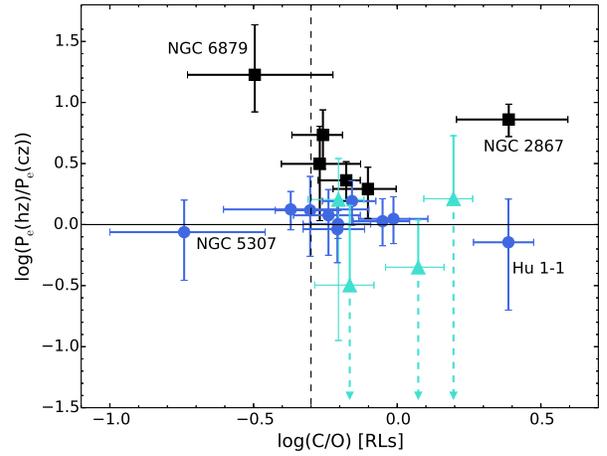}
\caption{Ratio between the pressure in the hot zones, $P_{\rm e}$(hz), and the pressure in the cold zones, $P_{\rm e}$(cz), 
as a function of $\log$(C/O) derived from RLs. The symbols and the solid line are the same as in Fig.~\ref{fig3}. 
The dashed line shows where C/O = 0.5, representative of the solar C/O value \citep{all02}. \label{fig6}}
\end{centering}
\end{figure}

\begin{figure}
\begin{centering}
\includegraphics[width = \hsize, trim = 20 20 40 20, clip = yes]{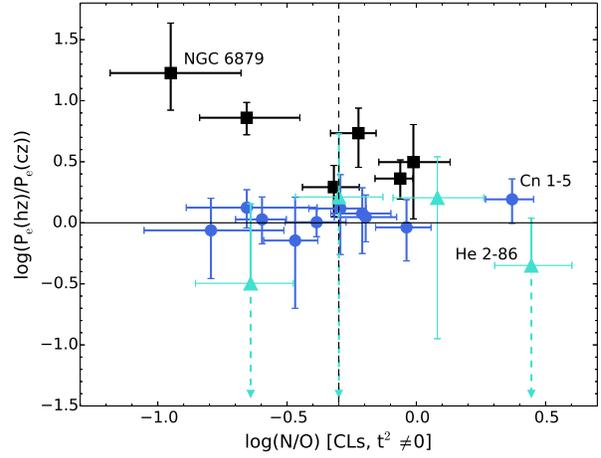}
\caption{Ratio between the pressure in the hot zones, $P_{\rm e}$(hz), and the pressure in the cold zones, $P_{\rm e}$(cz), 
as a function of the abundance 
ratios $\log$(N/O) derived from CLs for $t^2 \neq 0$. The symbols and the solid line are the same as in Fig.~\ref{fig3}. 
The dashed line shows where $\log$(N/O) = $-$0.3 which distinguish type I and type II PNe according to \citet{pem90}. 
\label{fig7}}
\end{centering}
\end{figure}

The C/O and N/O abundance ratios reflect the effect of nucleosynthesis mechanisms in the progenitor 
stars of PNe, the asymptotic giant branch stars (AGB). Theoretical models by \citet{kar10} predict an 
increase of the nitrogen abundance in the most massive AGB stars, above $\sim$4 $M_\odot$ for a metallicity of Z = 0.02, 
as a consequence of the second dredge up. These models predict that the carbon abundance is 
not altered in the less massive progenitors with $M < 2.5 M_\odot$, increases in stars with $M \gtrsim 2.5 M_\odot$ 
due to the third dredge up, and may decrease in stars with $M \gtrsim 4 M_\odot$ due to the hot bottom 
burning process. In Figures~\ref{fig6} and \ref{fig7} we studied possible correlations between pressure ratios and 
these abundance ratios.

The C/O abundance ratios were 
computed from RLs. The ionic abundances of C$^{++}$, shown in Table~\ref{Tabco}, were directly taken 
from the papers listed in Table~\ref{tab1}. In those PNe where no uncertainties are provided by the authors 
we assume a one sigma error of $\pm0.04$ dex for the determination of C$^{++}$/H$^+$. To calculate the 
total C/O abundance ratios, we use the ICFs and the associated uncertainties from \citet{del14}, shown in 
column (4) of Table~\ref{Tabco}. 

\begin{table}[!h]
\centering
\caption{Ionic abundances\tabnotemark{a}, ICF(C$^{++}$/O$^{++}$), and $\log$(C/O) values from RLs.}\label{Tabco}
\setlength{\tabnotewidth}{0.8\linewidth}
\setlength{\tabcolsep}{1.2\tabcolsep} 
\tablecols{7}
\begin{tabular}{lr@{$\pm$}lr@{$\pm$}lr@{}lr@{}l}
\toprule
\multicolumn{1}{c}{Object} & \multicolumn{2}{c}{\{C$^{++}$\}} & \multicolumn{2}{c}{ICF} & \multicolumn{2}{l}{$\log$(C/O)}\\ 
\midrule
Cn~1-5         & 9.08 & 0.02  & 0.95&$^{+0.05}_{-0.09}$ & 0.07&$_{-0.11}^{+0.09}$ \\
He~2-86       & 8.83 & 0.04  & 1.07&$^{+0.07}_{-0.09}$ & $-0.16$ &$_{-0.10}^{+0.08}$ \\
Hu~1-1         & 9.01 & 0.04  & 0.89&$^{+0.04}_{-0.09}$ & 0.39&$_{-0.12}^{+0.09}$ \\
Hu~2-1         & 8.66 & 0.04   & 0.81&$^{+0.03}_{-0.09}$ &$-0.16$&$_{-0.12}^{+0.08}$ \\
M~1-25         & 8.75 & 0.02  & 0.77&$^{+0.02}_{-0.09}$ &$-0.26$&$_{-0.11}^{+0.07}$ \\
M~1-30         & 8.93 & 0.01  & 0.64&$^{+0.03}_{-0.09}$ &$-0.18$&$_{-0.10}^{+0.05}$\\
M~1-61         & 8.64 & 0.03  & 1.06&$^{+0.07}_{-0.09}$ &$-0.20$&$_{-0.11}^{+0.09}$\\
M~3-15         & 8.84 & 0.04  & 1.14&$^{+0.08}_{-0.09}$ &$-0.27$&$_{-0.13}^{+0.14}$\\
NGC~2867   & 9.01 & 0.08   & 1.08&$^{+0.07}_{-0.09}$ & 0.39&$_{-0.18}^{+0.21}$ \\
NGC~5189   & 8.48 & 0.04   & 0.95&$^{+0.05}_{-0.09}$ &$-0.21$&$_{-0.12}^{+0.09}$\\
NGC~5307   & 7.95 & 0.10   & 1.20&$^{+0.26}_{-0.22}$ &$-0.74$&$_{-0.26}^{+0.28}$\\
NGC~6153   & 9.35 & 0.04   & 1.13&$^{+0.08}_{-0.09}$ &$-0.21$&$_{-0.12}^{+0.11}$ \\
NGC~6803   & 8.79 & 0.04   & 1.11&$^{+0.08}_{-0.09}$ &$-0.24$&$_{-0.12}^{+0.11}$ \\
NGC~6879   & 8.25 & 0.04   & 1.20&$^{+0.26}_{-0.22}$ &$-0.50$&$_{-0.23}^{+0.27}$\\
NGC~6891   & 8.74 & 0.04   & 1.17&$^{+0.09}_{-0.09}$ &$-0.01$&$_{-0.12}^{+0.12}$ \\
NGC~7009   & 8.74  & 0.04  & 1.18&$^{+0.26}_{-0.22}$ &$-0.37$&$_{-0.23}^{+0.27}$ \\
NGC~7026   & 8.98 & 0.04   & 1.04&$^{+0.06}_{-0.09}$ &$-0.09$&$_{-0.12}^{+0.10}$\\ 
PB~8            & 8.84 & 0.04   & 1.00&$^{+0.06}_{-0.09}$ &$-0.31$&$_{-0.12}^{+0.10}$ \\
PC~14          & 8.91 & 0.03   & 1.12&$^{+0.08}_{-0.09}$ &$-0.05$&$_{-0.10}^{+0.09}$\\
Pe~1-1         & 9.02 & 0.02   & 0.92&$^{+0.04}_{-0.09}$ &0.19&$_{-0.10}^{+0.07}$ \\  
\bottomrule
\tabnotetext{a}{\{X$^{+i}$\} = 12 + $\log$(X$^{+i}$/H$^+$).}
\end{tabular}
\end{table}

Using CLs and the physical conditions given in Table~\ref{Tt-oiii}, we computed the O$^{+}$, O$^{++}$, and 
N$^{+}$ abundances. We used the [\ion{O}{2}] $\lambda$3727, $\lambda$3729 lines to compute O$^{+}$/H$^+$, 
the [\ion{O}{3}] $\lambda$4959, $\lambda$5007 lines for O$^{++}$/H$^+$, and the [\ion{N}{2}] $\lambda$6548, 
$\lambda$6584 lines for N$^{+}$/H$^+$. The final ionic abundances are the mean values from those derived 
with each line ratio. The values of N/O have been derived with the ICFs by \citet{del14}. Columns 2, 4, and 6 
in Table~\ref{Tabno} list the ionic and total abundances derived from CLs and $t^2=0$, whereas columns 3, 5, 
and 7 show the values for $t^2\neq0$. 

Figures~\ref{fig6} and \ref{fig7} show that there is no obvious trend between the pressure ratios and the 
C/O or N/O abundance ratios. According to \citet{pem90} about half of our sample are type I PNe, with 
$\log$(N/O) $>$ 0.5, that would arise from massive progenitor stars, whereas the other half are type II 
PNe, with $\log$(N/O) $<$ 0.5. As for the C/O values, most of the PNe show C/O $>$ C/O$_\odot$ 
(where C/O$_\odot$ = 0.5; \citealt{all02}) probably reflecting the occurrence of the third dredge up, but 
not of the hot bottom burning. Only three 
PNe have C/O $<$ C/O$_\odot$: NGC~5307, NGC~6879, and NGC~7009. Since these three PNe 
also show low values of N/O, they probably arise from low mass progenitor stars ($M < 2 M_\odot$).

\begin{scriptsize}
\begin{table*}[!t]
\centering
\caption{Ionic abundances\tabnotemark{a}, ICF(N$^{+}$/O$^{+}$), and $\log$(N/O) values from CLs.} \label{Tabno}
\setlength{\tabnotewidth}{0.8\linewidth}
\setlength{\tabcolsep}{1.2\tabcolsep} 
\tablecols{15}
\begin{tabular}{lr@{}lr@{}lr@{}lr@{}lr@{}lr@{}lr@{}l}
\toprule
\multicolumn{1}{c}{Object} & \multicolumn{2}{c}{\{N$^{+}$\}} & \multicolumn{2}{c}{\{N$^{+}$\}} & \multicolumn{2}{c}{\{O$^{+}$\}} 
& \multicolumn{2}{c}{\{O$^{+}$\}} & \multicolumn{2}{c}{ICF} & \multicolumn{2}{c}{$\log$(N/O)} & \multicolumn{2}{c}{$\log$(N/O)}\\
\multicolumn{1}{c}{} & \multicolumn{2}{c}{$t^2=0$} & \multicolumn{2}{c}{$t^2\neq0$} & \multicolumn{2}{c}{$t^2=0$} & \multicolumn{2}{c}{$t^2\neq0$} 
& \multicolumn{2}{c}{} & \multicolumn{2}{c}{$t^2=0$} & \multicolumn{2}{c}{$t^2\neq0$}\\ 
\midrule
Cn~1-5         &  8.01 &$\pm0.03$ & 8.17 & $\pm0.03$ &  8.04 &  $\pm0.06$ &  8.25 &  $\pm0.06$ & 3.31 & $_{-0.26}^{+0.41}$ & 0.50 &  $_{-0.14}^{+0.16}$ &  0.44 &  $_{-0.14}^{+0.16}$ \\
He~2-86       &  7.58 &$\pm0.05$ & 7.76 & $\pm0.05$ &  7.73 &  $_{0.09}^{+0.11}$ &  7.97 &  $_{-0.09}^{+0.11}$ & 3.77 & $_{-0.29}^{+0.45}$ & 0.43 &  $_{-0.17}^{+0.18}$  &  0.37 &  $_{-0.17}^{+0.18}$ \\
Hu~1-1         &  7.53 &$\pm0.03$ & 7.68 &  $\pm0.03$ &  7.92 & $_{0.05}^{+0.04}$ &  8.13 & $_{-0.05}^{+0.04}$ & 0.95 & $_{-0.24}^{+0.38}$ & $-0.41$ &  $_{-0.12}^{+0.14}$ & $-0.47$ &  $_{-0.12}^{+0.14}$ \\
Hu~2-1         &  7.05 &$_{-0.04}^{+0.05}$ & 7.39 & $_{-0.04}^{+0.05}$ &  7.73 &  $_{-0.12}^{+0.21}$ &  8.17 &  $_{-0.12}^{+0.21}$ & 1.38 & $_{-0.21}^{+0.33}$ & $-0.53$ &  $_{-0.22}^{+0.17}$  & $-0.64$ &  $_{-0.22}^{+0.17}$ \\
M~1-25         &  7.99 &$\pm0.04$ & 8.15 &  $\pm0.04$ &  8.41 &  $\pm0.08$ &  8.62 &  $\pm0.08$ & 1.80 & $_{-0.20}^{+0.31}$ & $-0.16$ &  $_{-0.12}^{+0.14}$  & $-0.22$ &  $_{-0.12}^{+0.14}$ \\
M~1-30         &  8.33 &$\pm0.04$ & 8.56 &  $\pm0.04$ &  8.59 &  $_{-0.08}^{+0.09}$ &  8.90 &  $_{-0.08}^{+0.09}$ & 1.95 & $_{-0.14}^{+0.23}$ & 0.03 &  $_{-0.10}^{+0.11}$ & $-0.05$ &  $_{-0.10}^{+0.11}$ \\
M~1-61         &  7.23 &$\pm0.04$ & 7.38 &  $\pm0.04$ &  7.66 &  $_{-0.09}^{+0.10}$ &  7.87 &  $_{-0.09}^{+0.10}$ & 3.73 & $_{-0.29}^{+0.45}$ & 0.14 &  $_{-0.17}^{+0.18}$ &  0.08 &  $_{-0.17}^{+0.18}$ \\
M~3-15         &  6.92 &$\pm0.05$ & 7.13 &  $\pm0.05$ &  7.48 &  $_{-0.09}^{+0.08}$ &  7.76 &  $_{-0.09}^{+0.08}$ & 4.06 & $_{-0.12}^{+2.7}$ & 0.07 &  $_{-0.09}^{+0.57}$ & $-0.02$ &  $_{-0.09}^{+0.57}$ \\
NGC~2867   &  6.93 &$\pm0.03$ & 7.16 &  $\pm0.03$ &  7.44 &  $\pm0.06$ &  7.75 &  $\pm0.06$ & 0.86 & $_{-0.29}^{+0.45}$ & $-0.58$ &  $_{-0.16}^{+0.17}$ & $-0.66$ &  $_{-0.16}^{+0.17}$ \\
NGC~5189   &  7.85 &$_{-0.04}^{+0.03}$ & 8.01 &$_{-0.04}^{+0.03}$ &  7.76 &  $\pm0.06$ &  7.98 &  $_{-0.05}^{+0.06}$ & 0.86 & $_{-0.26}^{+0.41}$ & $0.03$ &  $_{-0.14}^{+0.15}$ & $-0.03$ &  $_{-0.14}^{+0.15}$ \\
NGC~5307   &  5.98 &$_{-0.02}^{+0.01}$ & 6.14 &$_{-0.02}^{+0.01}$ &  6.73 & $_{-0.05}^{+0.06}$  &  6.96 &  $_{-0.05}^{+0.06}$ & 1.06 & $_{-0.12}^{+2.7}$ & $-0.74$ &  $_{-0.08}^{+0.57}$ & $-0.80$ &  $_{-0.08}^{+0.57}$ \\
NGC~6153   &  7.19 &$\pm0.03$ & 7.61 &  $\pm0.03$ & 7.41 &  $\pm0.05$ &  8.01 &  $\pm0.05$ & 1.03 & $_{-0.30}^{+0.47}$ & $-0.22$ &  $_{-0.16}^{+0.17}$  & $-0.38$ &  $_{-0.16}^{+0.17}$ \\
NGC~6803   &  7.35 &$\pm0.03$ & 7.58 &  $\pm0.03$ &  7.55 &  $\pm0.06$ &  7.85 &  $\pm0.06$ & 1.17 & $_{-0.30}^{+0.47}$ & $-0.13$ &  $_{-0.16}^{+0.17}$   & $-0.21$ &  $_{-0.16}^{+0.17}$ \\
NGC~6879   &  5.62 &$\pm0.03$ & 5.94 &  $\pm0.03$ &  6.54 &  $_{-0.05}^{+0.06}$ &  6.97 &  $_{-0.05}^{+0.06}$ & 1.21 & $_{-0.12}^{+2.7}$ & $-0.84$ &  $_{-0.07}^{+0.57}$  & $-0.95$ &  $_{-0.07}^{+0.57}$ \\
NGC~6891   &  6.35 &$\pm0.03$ & 6.43 &  $\pm0.03$ &  7.14 &  $_{-0.05}^{+0.04}$ &  7.25 &  $_{-0.05}^{+0.04}$ & 4.16 & $_{-0.12}^{+2.7}$ & $-0.16$ &  $_{-0.06}^{+0.57}$  & $-0.20$ &  $_{-0.06}^{+0.57}$ \\
NGC~7009   &  6.54 &$\pm0.02$ & 6.85 &  $\pm0.02$ &  7.09 &  $_{-0.03}^{+0.04}$ &  7.50 &  $_{-0.03}^{+0.04}$ & 0.98 & $_{-0.12}^{+2.7}$ & $-0.55$ &  $_{-0.06}^{+0.57}$  & $-0.66$ &  $_{-0.06}^{+0.57}$ \\
NGC~7026   &  7.63 &$_{-0.03}^{+0.04}$ & 7.88 &  $_{-0.03}^{+0.04}$ &  7.85 &  $_{-0.06}^{+0.14}$ &  8.19 &  $_{-0.06}^{+0.14}$  & 0.97 & $_{-0.27}^{+0.43}$ & $-0.23$ &  $_{-0.19}^{+0.16}$ & $-0.32$ &  $_{-0.19}^{+0.16}$ \\ 
PB~8            &  7.16 &$\pm0.03$ & 7.39 &  $\pm0.03$ &  7.92 &  $\pm0.07$ &  8.23 &  $\pm0.07$ & 3.59 & $_{-0.28}^{+0.43}$ & $-0.20$ &  $_{-0.15}^{+0.16}$ & $-0.29$ &  $_{-0.15}^{+0.16}$ \\
PC~14          &  6.88 &$\pm0.03$ & 7.05 &  $\pm0.03$ &  7.49 &  $_{-0.05}^{+0.06}$ &  7.72 &  $_{-0.05}^{+0.06}$ & 1.18 & $_{-0.30}^{+0.47}$ & $-0.53$ &  $_{-0.16}^{+0.17}$   & $-0.60$ &  $_{-0.16}^{+0.17}$ \\
Pe~1-1         &  7.38 &$_{-0.06}^{+0.09}$ & 7.53 &  $_{-0.06}^{+0.09}$ &  7.93 &  $_{-0.14}^{+0.22}$ &  8.14 &  $_{-0.14}^{+0.22}$ & 1.99 & $_{-0.26}^{+0.40}$ & $-0.25$ &  $_{-0.21}^{+0.19}$  & $-0.31$ &  $_{-0.21}^{+0.19}$ \\  
\bottomrule
\tabnotetext{a}{\{X$^{+i}$\} = 12 + $\log$(X$^{+i}$/H$^+$).}
\end{tabular}
\end{table*}
\end{scriptsize}

\section{High density objects}

There are four objects where we can not rule out the presence of high density
low temperature regions, they  are Cn~1-5, Hu~2-1, M~1-61 and Pe~1-1. Three of these
objects, Hu~2-1, M1-61 and Pe 1-1, are of relatively high density, where our method
to derive \dens(\ion{O}{2}) is not very sensitive and presents large errors because 
it saturates at high densities (see Figure 3 of PP13). Cn~1-5 is of relatively low 
density and it is possible to increase the accuracy of the \dens(\ion{O}{2}) determination 
with additional observations.

Of these four objects only one, Hu~2-1 has a high ADF of 5.05. Alternatively, 
the other three have relatively low ADF values: 2.08 for Cn~1-5, 1.80 for M~1-61, 
and 1.81 for Pe~1-1.
These values are smaller than the ADF average of the other sixteen objects
that amounts to 2.75, and we consider unlikely the presence of high density
knots of low temperature in these three objects.

\section{NGC 6153}\label{SSO/Hrich}

Of our sample of 20 PNe the one with the highest ADF values is NGC 6153. This object 
shows $T(Bac)$ = 6080 K and $T(\ion{O}{ii})$ = 5780 K, values that are very similar and considerably
smaller than $T[\ion{O}{3}]$ = 9120 K, (see Tables~\ref{Tt-oii} and \ref{Tt-oiii}), in addition the large variations
in the ratio of the 4363 to F1 [\ion{O}{3}] line intensities as a function of velocity indicate
the presence of very large temperature variations within the nebula \citep{bar06}.

The densities and temperatures derived from the recombination lines 
of \ion{O}{ii} rule out the models with densities of $2\times 10^6$ ${\rm cm}^{-3}$, 
and also the models with H depleted material with temperatures of about 
500 K discussed by \citet{liu00}. 

Moreover, \citet{peq02} have presented a two phase photoionization model for this object
where component one, C1, has a \temp$\sim$10$^3$ K and an \dens$\sim4\times10^4$ cm$^{-3}$ and
component two, C2, has a \temp$\sim6\times10^3$ K and an \dens$\sim5\times10^3$ cm$^{-3}$,
the gas pressures in C1 and C2 are generally found within a factor of two of each other
approximately in pressure equilibrium, the \ion{O}{2} temperatures and densities derived by
us for NGC 6153 from observations, see Table 2, are in disagreement with this
model and therefore rule it out.

Similarly \citet{yua11} present a 2 phase model for NGC~6153 that approximately represents 
many of the observed lines. In this model the emission of the \ion{O}{2} lines comes mostly 
from metal rich inclusions with \temp = 815 K and \dens = 6680 cm$^{-3}$. While this model 
does a reasonable job at reproducing most of the observed line intensities, it fails to reproduce 
the observed $I$(4649)/$I$(VI) ratio: a) the ratio for the model presented in the paper is 0.400, 
the high density limit (probably the model does not include the non-LTE atomic physics required 
by these densities); b) even if it were considered, the expected line intensity ratio 
$I$(4649)/$I$(4638+51+61) = 1.189 for the sum of the components of the bi-abundance model, 
while the observed ratios by \citep{liu00} is 1.058  (Liu et al do not present errors, but by studying 
other faint line with known intensity ratios, we estimate the error in this ratio to be approximately 
4.5\%). The electron density required to produce this line ratio, at 815 K, is 
\dens = $2200^{+600}_{-400}$ cm$^{-3}$.

A necessary modification to \citet{yua11} model, to maintain their abundances and temperatures, is to consider 
metal rich inclusions 3 times less dense and 3 times more massive. This model would have the 
capacity to reproduce the observed \ion{O}{2} V1 as well as the observed [\ion{O}{3}] lines. In 
this model, most of the oxygen would be on the metal rich inclusions. 
However the observed intensity of \ion{O}{2} $\lambda$4649 has two unfortunate consequences 
for this model: a) the electron density of the metal rich inclusions would be lower than the one 
from the ambient medium and b) the pressure of the metal rich inclusions would be about a factor 
of 20 lower than the one of the ambient medium; it is difficult to imagine a scenario that would 
allow such inclusions to be formed and to survive without being mixed or compressed by the 
ambient medium.

Overall, a bi-abundance scenario where the abundance of the cold metal rich inclusions 
have high density is ruled out by the observed OII V1 line ratios, and a scenario where the 
cold metal rich inclusions have low density has to be thought carefully before being considered.

This discussion could be extended for most objects of our sample (at least 16).
Difficulties with bi-abundance models will be larger for objects where the measured 
\dens(\ion{O}{2}) is smaller than the \dens([\ion{Cl}{3}]).

\section{Conclusions}\label{Scon}

We have studied a sample of 20 PNe that have been observed with
high spectral resolution and high quality line intensity
determinations. The main conclusions of our work follow.

1) We have found that the determination of the \dens(\ion{O}{2}) values 
is a very important tool to test the existence of objects with metal-rich 
inclusions embedded in a lower density medium.

2) In sixteen of the objects the \ion{O}{ii} lines originate
in low density regions, or in other words they do not have high
density clumps producing most of the \ion{O}{ii} line intensities.

3) {For four objects of the sample, Cn 1-5, Hu 2-1, M 1-61, and Pe 1-1, 
the $I$(4649)/$I$(4639+51+62) ratio error bars reach the high density limit 
and therefore, we were not able to obtain reliable values of the densities 
where the \ion{O}{ii} lines originate.}

4) There are two results that indicate that in the PNe of
the sample H, He and O are well mixed: a) For a subsample of 19 
objects the average \temp(V1/F1) amounts to  7610 K, 
similar to the average \temp(Bac/Pac) that amounts to 8030 K. b) For a
subsample of 12 objects the average $t^2$(O$^{++}$) amounts
to 0.045, similar to the average $t^2$(He$^{+}$/CL) that amounts
to 0.042.

5) The observed $t^2$(O$^{++}$) values, that are in the 0.024 to 0.128 
range, are considerably higher than the predicted $t^2$(O$^{++}$) values 
by photoionization models, that are lower than 0.012. 
This result implies that in addition to photoionization
other sources of energy are needed to explain the observed
$t^2$(O$^{++}$) values. For the objects where the pressure
from CLs is higher than the pressure from RLs, we suggest that
shock waves could be the main cause for the high $t^2$(O$^{++}$)
observed values.

6) For many PNe of low density we find $P_{\rm e}$(CLs)/$P_{\rm e}$({\rm RLs}) closer
to 1 than in PNe of high density, a similar result is found in H II
regions. Presumably the PNe and H II regions with lower densities
are older and presumably the effect of shocks becomes smaller
as a function of time.

7) For most of the PNe of our sample we find that the \dens(\ion{O}{2}) 
values are {similar} than the \dens([\ion{Cl}{3}]) ones. 
In addition we find that the volume where the bulk of the O II RLs 
originate is {similar} than the volume where the bulk of the 
[O III] lines originate. These two results are contrary to the models 
that postulate the presence of pockets of high density and low 
temperature embedded in a medium of lower density and higher temperature.

8) Based on the $I$(4649)/$I$(4639+51+61) ratio of multiplet V1 of \ion{O}{2} 
(see Figure 3 of PP13) we find that the use of lower temperatures than the ones 
adopted in this paper for the \ion{O}{2} zone implies lower \dens(\ion{O}{2}) values 
than the ones presented in Table 2. This would strengthen conclusion 7.\\

{We thank the referee for a critical reading of the manuscript and several useful 
suggestions.}
G. Delgado$-$Inglada gratefully acknowledges a DGAPA postdoctoral grant from 
the Universidad Nacional Aut\'onoma de M\'exico (UNAM). J. Garc\'ia-Rojas 
acknowledges funding by the Spanish Ministry of Economy and Competitiveness 
(MINECO) under the grant AYA2011-22614 and Severo Ochoa SEV-2011-0187. 
A. Peimbert, M. Peimbert, and M. Pe\~na are grateful for the finantial support 
provided by CONACyT (grant 129753).

\end{document}